\documentclass[twocolumn,showpacs,pra]{revtex4}
\usepackage{graphicx}
\usepackage{dcolumn}
\usepackage{bm}

\begin{document}

\preprint{APS/123-QED}

\title{Solitons in a trapped spin-1 atomic condensate}
\author{Wenxian Zhang,$^{1,3}$ \"O. E. M\"ustecapl{\i}o\u{g}lu,$^{2}$ and L.
You$^{3}$}

\affiliation{$^{1}$Ames Laboratory, Iowa State University, Ames,
Iowa 50011, USA}
\affiliation{$^2$Department of Physics, Ko\c{c}
University, Sar{\i}yer, Istanbul, 34450, Turkey,}
\affiliation{$^3$School of Physics, Georgia Institute of
Technology, Atlanta, Georgia 30332, USA}

\date{ \today }

\begin{abstract}
We numerically investigate a particular type of spin solitons
inside a trapped atomic spin-1 Bose-Einstein condensate (BEC)
with ferromagnetic interactions. Within the mean field theory
approximation, our study of the solitonic dynamics shows that
the solitonic
wave function, its center of mass motion, and the local spin
evolutions are stable and are intimately related to the domain
structures studied recently in spin-1 $^{87}$Rb condensates. We
discuss a rotating reference frame wherein the dynamics of the
solitonic local spatial spin distribution become time independent.
\end{abstract}


\maketitle

\section{Introduction}
\label{sec:intr}

Solitonic structure is an interesting and common subject in
nonlinear science. It has been extensively studied in nonlinear
optics \cite{AgrawalBook89}, fluids \cite{DrazinBook89}
and plasmas \cite{Lonngren83}, magnetic films \cite{Wu04},
nanotubes \cite{Lerblond04}, and recently in quantum
superfluid of atomic Bose-Einstein
condensates \cite{Strecker02}. In nontechnical terms, the shape
of a soliton remains unaltered during propagation due to a
balance of the spreading from its dispersion with nonlinear
interactions. This feature has been applied with great success
in optical soliton based advanced communications.

As realized in recent years an atomic Bose-Einstein condensate (BEC)
provides a new and powerful tool for the study of nonlinear
phenomena. At weak interactions, a condensate is successfully
modelled by the Gross-Pitaevskii (GP) equation, or the nonlinear
Schr\"odinger equation (NLSE)
\cite{Anderson95, Davis95, Bradley95, PethickBook02,
PitaevskiiBook03, Dalfovo99}. Feshbach resonance in recent years
has provided a practical mechanism for adjusting the atomic
s-wave scattering length through which both the sign and
the strength of the atomic interactions become adjustable \cite{Cornish00}.
Bright matter wave soliton trains were
realized in a $^7$Li condensate due to its attractive atom-atom interaction,
while gray and dark solitons were realized for repulsively interacting
condensates through the phase engineering technique on the
condensate wave function \cite{Burger99,Denschlag00}.

Solitons with more than one parameter are often referred to as vector solitons.
For instance, vector solitons are created in nonlinear optics if different
polarizations or spatial modes of light in an optical fiber are taken as
parameters. This could easily be implemented in a spinor condensate as the
atomic internal degrees of freedom become available. Many have suggested
creating a Manakov soliton in a two-component BEC with attractive
self-interactions \cite{Babarro05}. Symbiotic solitons have also been
investigated in a two-component BEC with repulsive self-interactions
accompanied by attractive cross-interactions \cite{Perez-Garcia05}. Domain wall
solitons can be created in a trapped two-component condensate even when both
self- and cross-interactions are repulsive \cite{Coen01}. More generally,
self-consistent collective excitations of an atomic condensate, or higher
eigen-states of the Gross-Pitaevskii equation, are also called solitons loosely
because of their reasonable stability and resemblance to other properties of
solitonic states \cite{au}.

With a far off resonant optical trap, all hyperfine states of an atom can be
confined simultaneously, thus realizing a spinor condensate, as has been
reported in many experiments with $^{23}$Na and $^{87}$Rb atoms
\cite{Stenger98, Stamper-Kurn98, Stamper-Kurn99, Barrett01, Chang04,
Schmaljohann04, Kuwamoto04, Higbie05, Chang05}. The releasing of the spin
degree of freedom has led to many novel properties lacking in a single
component or scalar condensate. For instance, the spin exchange interaction as
in a spin-1 condensate, whereby two $|F=1,m_F=0\rangle$ atoms are converted
into one $|F=1,m_F=+1\rangle$ atom, and one $|F=1,m_F=-1\rangle$ atom upon an
elastic collision, already well understood \cite{Ho98, Ohmi98, Law98}, also
helps to support the existence of spinor solitons. In a homogeneous system,
exact soliton solutions for an attractively interacting spin-1 condensate have
been recently obtained analytically \cite{Ieda04}. Also theoretical
investigations have been carried out on magnetic (or spinor) solitons in a
ferromagnetically interacting spin-1 condensate \cite{Xie04,Li05}.

In this paper, we focus on the
more realistic case of spin solitons inside trapped atomic
condensates (of $^{87}$Rb). First we attempt to numerically
find solitonic structures through the propagation in time of the
GP equation for a trapped spin-1 condensate. Then we
demonstrate that the dynamics of the found spin solitons is stable
and is related to the multi-domain structure as studied recently
\cite{Zhang05b,Saito05}. Finally, we discuss a rotating reference
frame wherein the local spin distribution of the spinor condensate
becomes stationary like a spatially propagating soliton in a
moving reference frame.

This paper is organized as follows. In Sec. \ref{sec:intr}
we give an introduction and discuss the motivations for our work.
The theoretical formulation is based on a model system of a spin-1
condensate as is
presented in Sec. \ref{sec:npse}. Section \ref{sec:ss} describes
the numerically obtained soliton state, while Sec. \ref{sec:tp}
focuses on the investigation of the solitonic state's dynamics,
center of the mass motion, and local spin distribution.
In Sec. \ref{sec:sr} we discuss a rotating reference frame and
show that the local spin distribution becomes
time-independent within this frame.
Finally we conclude in Sec. \ref{sec:cl}.

\section{Review of an effective quasi-one-dimensional description}
\label{sec:npse}

A gas of interacting spin-1 atoms is described by the following
second quantized Hamiltonian (summation over repeated indices is
assumed) \cite{Ho98, Ohmi98, Law98}
\begin{eqnarray}
H &=&\int d\vec r\, \left[ \Psi_i^\dag\left(-{\hbar^2 \over 2M}\nabla^2
+ V_{\rm ext}(\vec r)\right) \Psi_i \right. \nonumber\\
&&+\left. {\frac{c_0}{2}} \Psi_i^\dag\Psi_j^\dag\Psi_j\Psi_i +{c_2\over 2}
\Psi_k^\dag\Psi_i^\dag\left(F_\gamma
\right)_{ij}\left(F_\gamma\right)_{kl}\Psi_j \Psi_l\right], \hskip 14pt
\label{eq:h}
\end{eqnarray}
where $\Psi_j(\vec r)$ ($\Psi_j^\dag$) is the field operator that annihilates
(creates) an atom in the $j$th internal state at location $\vec r$, $j=+,0,-$
denotes atomic hyperfine state $|F=1, m_F=+1,0,-1\rangle$, respectively. $M$ is
the mass of each atom, and $V_{{\rm ext}}(\vec r)$ is an internal state
independent trap potential. Terms with coefficients $c_0$ and $c_2$ of Eq.
(\ref{eq:h}) describe elastic collisions of two spin-1 atoms expressed in
terms of the scattering lengths $a_0$ ($a_2$) in the combined symmetric channel
of total spin $0$ ($2$), $c_0=4\pi\hbar^2(a_0+2a_2)/3M$, and
$c_2=4\pi\hbar^2(a_2-a_0)/3M$. $F_{\gamma=x,y,z}$ are spin-1 matrices.

The equation of motion for the field operator in the Heisenberg picture is
given by
\begin{eqnarray}
i\hbar \frac {\partial}{\partial t}\Psi_i(\vec r, t) &=& \left[\Psi_i,
H\right].
\end{eqnarray}
Adopting a mean field theory approach by assuming that
the condensate consists of a large number of atoms,
we introduce the condensate order parameter or wave function
$\Phi_i=\langle\Psi_i\rangle$ for the $i$th component. Neglecting quantum
fluctuations we arrive at the coupled Gross-Pitaevskii equations
\begin{eqnarray}
i\hbar {\partial \over \partial t}\Phi_+ &=&
    \left[{\cal L}
    + c_2(n_++n_0-n_-)\right]\Phi_++c_2\Phi_0^2\Phi_-^*,\nonumber\\
i\hbar {\partial \over \partial t}\Phi_0 &=&
    \left[{\cal L}
    + c_2(n_++n_-)\right]\Phi_0+2c_2\Phi_+\Phi_-\Phi_0^*,\label{eq:gpe}\\
i\hbar {\partial \over \partial t}\Phi_- &=&
    \left[{\cal L}
    + c_2(n_-+n_0-n_+)\right]\Phi_-+c_2\Phi_0^2\Phi_+^*,\nonumber
\end{eqnarray}
where ${\cal L} = -(\hbar^2/2M)\nabla^2+V_{\rm ext}(\vec r) + c_0n $, $n =
\sum_i n_i$ is the total condensate density, and $n_i = |\Phi_i|^2$. Another way
to derive the above coupled GP Eqs. (\ref{eq:gpe}) is to take the variations of
the energy functional with respect to the condensate wave function
\begin{eqnarray}
i\hbar\,{\partial \over \partial t}\Phi_i &=& {\delta E[\Phi_i, \Phi_i^*]\over
\delta \Phi_i^*}, \nonumber
\end{eqnarray}
where
\begin{eqnarray}
E[\Phi_i, \Phi_i^*] &=&\int d\vec r\, \left[ \Phi_i^*\left(-{\hbar^2 \over
2M}\nabla^2 + V_{\rm ext}+{1\over 2}c_0n\right)
\Phi_i \right. \nonumber \\
&&+ \left.{1\over 2}c_2 \Phi_k^*\Phi_i^*\left(F_\gamma
\right)_{ij}\left(F_\gamma\right)_{kl} \Phi_j \Phi_l\right].
\label{eq:e}
\end{eqnarray}

Inside a trap with tight radial confinement, the condensate
assumes a prolate shape. Several effective one-dimensional (1D)
approaches have already been developed \cite{Jackson98,
Salasnich02, Salasnich04, Chiofalo00, Gerbier04, Zhang05}, with
the simplest of them assuming a fixed transverse Gaussian profile.
Recent studies, however, have indicated that the effective
quasi-1D non-polynomial Schr\"odinger equation (NPSE) is the more
appropriate choice. In fact for most systems of interest, NPSE
represents the most powerful and efficient tool especially in the
weakly interacting limit \cite{Jackson98, Salasnich02,
Salasnich04, Zhang05}. In several recent experiments, a single
running wave optical trap is used to confine spin-1 atomic
condensates, a situation well described by the
quasi-one-dimensional trapping geometry \cite{Schmaljohann04,
Kuwamoto04, Higbie05, Chang04}. More precisely the external
cigar-shaped trap is described by a harmonic trap potential
$V_{\rm ext} = (M/2)(\omega_\perp^2 r_\perp^2 + \omega_z^2 z^2)$
with $\omega_x=\omega_y =\omega_\perp$, $r_\perp=\sqrt{x^2+y^2}$,
and $\omega_z \ll \omega_\perp$. The quasi-one-dimensional NPSE
description assumes a factorized wave function
\begin{eqnarray}
\Phi_i(\vec r_\perp,z;t) &=& \sqrt N \,\phi_\perp(\vec r_\perp; \chi(z,t))
\phi_i(z,t),
\end{eqnarray}
into the transversal and longitudinal functions \cite{Zhang05},
which in the case of a spin-1 condensate are governed respectively, by
\begin{eqnarray}
i\hbar {\partial \over \partial t}\phi_+ &=&
    \left[h_0 + c_2N\eta (\rho_++\rho_0-\rho_-)\right]\phi_+
    + c_2N\eta \phi_0^2\phi_-^*,\nonumber\\
i\hbar {\partial \over \partial t}\phi_0 &=&
    \left[h_0 + c_2N\eta(\rho_++\rho_-)\right]\phi_0
    + 2c_2N\eta \phi_+\phi_-\phi_0^*,\nonumber\\
i\hbar {\partial \over \partial t}\phi_- &=&
    \left[h_0 + c_2N\eta(\rho_-+\rho_0 -
    \rho_+)\right]\phi_- + c_2N\eta \phi_0^2\phi_+^*, \nonumber \\
\rho {\partial E_\perp \over \partial \chi} &+& \left({c_0N\over 2}\rho^2 +
{c_2N\over 2}S_2 \right){\partial \eta \over \partial \chi} = 0.
\label{eq:npse}
\end{eqnarray}
In the above, $N$ is the total number of condensed atoms. $\phi_i$ is the
factorized longitudinal envelop function of the quasi-one-dimensional wave function,
and it depends only on $z$ and $t$.
$\phi_\perp$ is the transversal wave function, satisfying $\int d\vec r_\perp
|\phi_\perp|^2 = 1$, and is assumed to be identical for all three spin components.
$\chi$ is a variational functional characterizing the width of the transversal wave
function. $h_0 = -(\hbar^2 / 2M)(\partial^2 / \partial z^2)+ V(z) + E_\perp +
c_0N \eta\rho$ with $V(z) = (M/2)\omega_z^2z^2$, and $E_\perp$ is the
transverse mode energy functional,
\begin{eqnarray}
E_\perp(\chi) &=& \int d\vec r_\perp \phi_\perp^*\left[-{\hbar^2\over 2M}\nabla_\perp^2
    + {1\over 2}M\omega_\perp^2 r_\perp^2\right]\phi_\perp.
\end{eqnarray}
$\eta(\chi)
= \int d\vec r_\perp |\phi_\perp|^4$ is a scaling factor for the nonlinear interaction strength.
$\rho(z) = \sum_i|\phi_i|^2$, and $S_2$
is independent of $\chi$ and is shown to be given by
\begin{eqnarray}
S_2 &=& |\phi_+|^4+|\phi_-|^4+2|\phi_+|^2|\phi_0|^2
    +2|\phi_-|^2|\phi_0|^2 \nonumber \\
    && - 2|\phi_+|^2|\phi_-|^2 + 2\phi_0^{*2}\phi_+\phi_-
    + 2\phi_+^*\phi_-^*\phi_0^2.
\end{eqnarray}
In obtaining the above relations self-consistently,
we have also assumed a weak time and $z$ dependence of the transverse wave
function, i.e., $\partial \phi_\perp /\partial t \simeq 0$ and $\nabla^2
\phi_\perp \simeq \nabla_\perp^2\phi_\perp$.

For a condensate with a large number of atoms, as in most current experiments,
the density distribution of the transversal direction approaches the
Thomas-Fermi (TF) limit, $\mu \gg \hbar \omega_\perp$, thus we take the TF
ansatz for the transverse wave function,
\begin{eqnarray}
\phi_\perp(\vec r_\perp; \chi(z,t)) &=& \left\{\begin{array}{cc}\sqrt{2\over
\pi}\; {1\over \chi}\; \sqrt{1-\left({r_\perp \over
\chi}\right)^2},&r_\perp\leq \chi;
\\0,&r_\perp > \chi. \end{array}\right. \nonumber
\end{eqnarray}
The kinetic energy in the transverse direction is therefore neglected
under the TF limit, leading to the transverse mode energy and
scaling factor being
\begin{eqnarray}
E_\perp
&=& {\hbar\omega_\perp\over 6}
\left({\chi^2\over a_\perp^2}\right), \nonumber\\
\eta &=& {4\over 3\pi \chi^2}. \nonumber
\end{eqnarray}

\section{Solitons in a trapped spin-1 Bose condensate}
\label{sec:ss}

Condensate solitons we discuss in this paper are simply self-consistent high
energy eigen-states of the GP Eq. (\ref{eq:gpe}) \cite{au}. Armed with the
quasi-1D NPSE approach and the imaginary time propagation method, we devise the
following operational procedure for an extensive search of the solitonic
states. Numerically we first find the ground state, then we propagate the
quasi-1D NPSE Eq. (\ref{eq:npse}) in the imaginary time domain \cite{Yi02},
continuously projecting out the ground state at each time step during
evolution,
\begin{eqnarray}
|\phi^{(n)}\rangle &\rightarrow & |\phi^{(n)}\rangle-\langle \phi^{(0)} |
\phi^{(n)}\rangle |\phi^{(0)}\rangle,
\end{eqnarray}
hoping for a self-consistent eigen-solution to be found. It is well known that
different eigen-solutions of a nonlinear Hamiltonian may not be orthogonal,
furthermore the family of all eigen-solutions does not generally constitutes an
orthornormal and complete basis. Yet, our procedure is found to converge at
least for the first self-consistent collective excitation. This solitonic state
we find is essentially independent of the choice for the transverse mode as
being a fixed Gaussian or a fixed TF ansatz.

\begin{figure}
\includegraphics[width=3.25in]{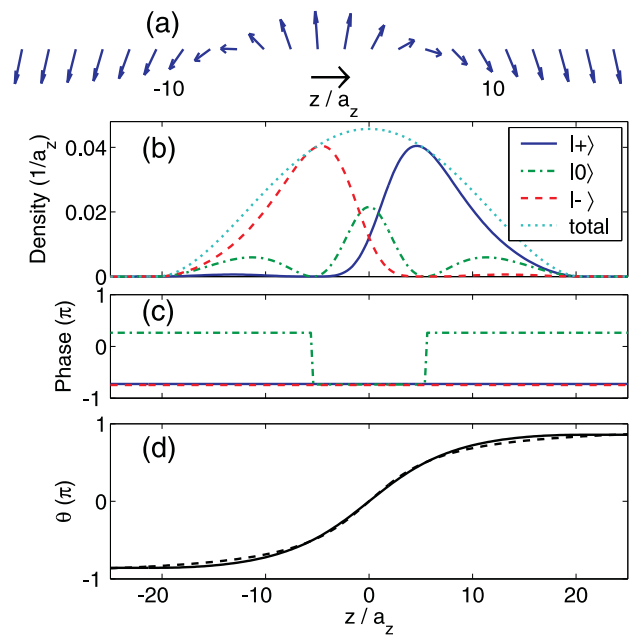}
\caption{(Color online) The spatial distribution of the local spin (a), the
density (b), and the phase (c) of a soliton state in a trapped spin-1 $^{87}$Rb
condensate. The trap parameters are $\omega_x=\omega_y=(2\pi) 240$ Hz, and
$\omega_z=(2\pi) 24$ Hz. The total number of atoms in the trap is $N=10^6$, and
the total magnetization (${\cal M}=N_+-N_-$) is zero. The solid curves in
panels (b) and (c) denote the density and the phase for the $|+\rangle$
component, the dash-dotted curves denote the $|0\rangle$ component, the dashed
curves denote the $|-\rangle$ component, and the dotted curve in (b) denotes
the total density. The solid line in panel (d) shows the local spin angle
distribution and the dashed line is that for a magnetic kink soliton in a homogeneous
system.}
\label{fig:s} 
\end{figure}

An example of the spin-1 soliton found in a trapped $^{87}$Rb condensate is
illustrated in Fig. \ref{fig:s}. The scattering lengths used are $a_0=101.8$ $a_B$
and $a_2=100.4$ $a_B$ with $a_B$ being the Bohr radius \cite{para2}.
In the subplot Fig. \ref{fig:s}(a) we display the spatial distribution
of the local spin average that is defined as
\begin{eqnarray}
f_x (z) &=& \langle F_x\rangle
    = \sqrt 2 \;{\rm Re} {[\phi_0(\phi_+^* + \phi_-^*)]}/\rho, \nonumber \\
f_y(z) &=& \langle F_y\rangle
    = -\sqrt 2\; {\rm Im} {[\phi_0^*(\phi_+ - \phi_-)]}/\rho, \nonumber \\
f_z(z) &=& \langle F_z\rangle
    = (|\phi_+|^2-|\phi_-|^2)/\rho,
\label{eq:sd}   
\end{eqnarray}
whose magnitude is given by $f(z)=\sqrt{f_x^2+f_y^2+f_z^2}$.
We find that the component $f_y$, which is displayed perpendicular to the plane,
remains very close to zero for the soliton, i.e., the local spins
stay almost in the $x$-$z$ plane. We believe the residue small but nonzero value of
$f_y$ is due to the numerical precision. Instead of getting rid of these
annoying numerical errors, we keep them to study the dynamical stability of the
soliton state. Figure \ref{fig:s}(a) clearly shows that the local spin rotates
nearly a complete revolution, or $2\pi$ around the $y$-axis from one side to
the other side of the condensate along the quasi-1D axial direction.
It is easy to check that the winding number of this solitonic spin
distribution is about one along the $y$-axis, which shows that the soliton state
we find might be related to a quasi-one-dimensional vortex.

We illustrate also the density and the phase distribution in Fig.
\ref{fig:s}(b) and \ref{fig:s}(c). We see a $\pi$ phase shift in the
$|0\rangle$ component near the central region where the density of the
$|0\rangle$ component diminishes. Also we notice that the $|+\rangle$ and
$|-\rangle$ components are symmetric with respect to the trap center at $z=0$.
The panel (d) shows the angular distribution of the local spin (solid line),
$\cos\theta = f_z/f$. It is compared to the soliton angular distribution
(dashed line) of a magnetic kink soliton given by $\theta = 2 \arctan(z/\xi)$
where $\xi=5.4a_z$ and the phase of $|0\rangle$ component jumps in its
neighborhood. The good agreement indicates that the soliton state we find in a
trapped spin-1 condensate is closely related to a magnetic kink soliton.

\section{Real time Propagation of the soliton in a trap}
\label{sec:tp}

Starting from the spinor soliton state found above, we can propagate in real
time the quasi-1D NPSE in the trap. This allows us to easily check the dynamic
stability of the solitonic structure. In addition, we can check several
conservative quantities such as the energy, the total number of atoms $N$, and
the total magnetization ${\cal M}=N_+-N_-$ to confirm the accuracy of our
numerical integration. During the propagation time $t\in [0,5000]$
($1/\omega_z$), the relative fluctuations of the energy, the number, and the
total magnetization are found to be less than $10^{-6}$, $10^{-6}$, and
$10^{-12}$, respectively.

\begin{figure}
\begin{center}
\includegraphics[width=3.35in]{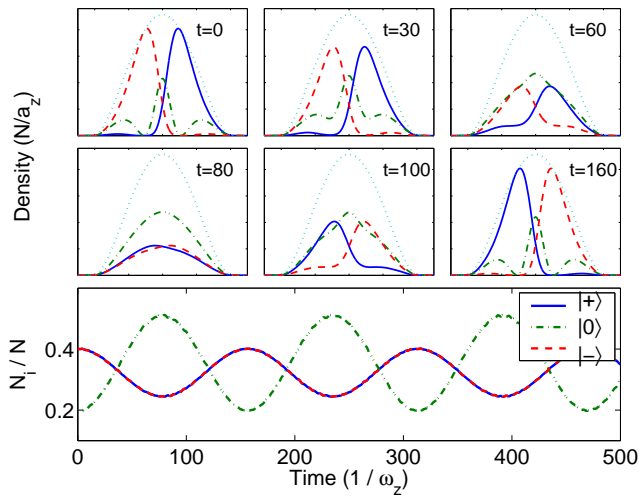}
\caption{(Color online) Temporal propagation of the spatial profile
of the trapped soliton. The parameters and notations are the same as in Fig.
\ref{fig:s}. The upper and the
middle rows display the density distributions of the soliton at different times.
The lower row shows the time dependence of the fractional populations of every
spin components.}
\label{fig:ts} 
\end{center}
\end{figure}

We show in Fig. \ref{fig:ts} the time dependence of the fractional population
for each component, clearly periodic in time. The period is found to be $T\sim
313$ $(1/\omega_z)$, or equivalent to about $2.1$ s in real time unit. For a
homogeneous condensate this period approaches infinity in the thermal dynamic
limit, a topic that is beyond the scope of the present article and also
currently unreachable in experiments. For a finite sized condensate as
discussed here, the period is found to decrease. Figure \ref{fig:ts} also shows
typical density distribution for each component at various times during the
first half period. We see the $|+\rangle$ and $|-\rangle$ components tunnel
through each other with the assistance of the $|0\rangle$ component and return
to their original locations within each period. We note that the $|+\rangle$
and $|-\rangle$ components are immiscible and the tunnelling is generally
prohibited without the presence of the $|0\rangle$ component. We do not see
such an immiscibility here, since the $|0\rangle$ component provides a cohesion
for the $|+\rangle$ and the $|-\rangle$ components. This type of tunnelling
process can also be explained in terms of a mutual precession among the three
components as in Ref. \cite{Ieda04}. We actually extract the precession angles
for both the $|+\rangle$ and $|-\rangle$ components and both are found equal to
$\pi$, corresponding to a simple exchange of their positions after the
tunnelling, i.e., spin up become spin down and vice versa. The interesting
feature for a trapped system, distinct from the homogeneous case, is that the
tunnelling process repeats itself cyclically upon reflections from the trap
boundaries.

\begin{figure}
\begin{center}
\includegraphics[width=3.35in]{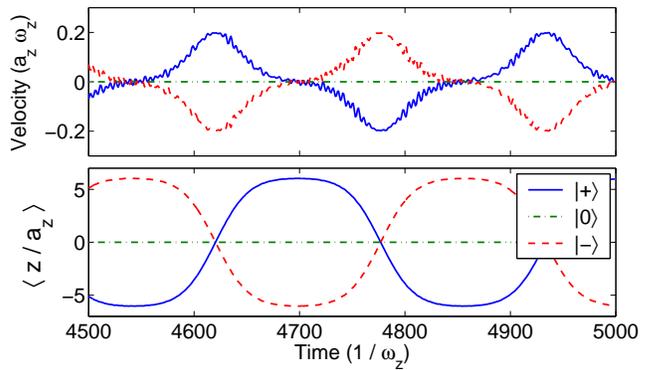}
\caption{(Color online) The velocity and the position for the center of mass
of each soliton component propagating in a trap. The parameters and
notations are the same as in Fig. \ref{fig:s}. The motion is
periodic, and no noticeable distortion occurs after more than 15
periods or about 33 s.}
\label{fig:tszv} 
\end{center}
\end{figure}

\begin{figure}
\begin{center}
\includegraphics[width=3.45in]{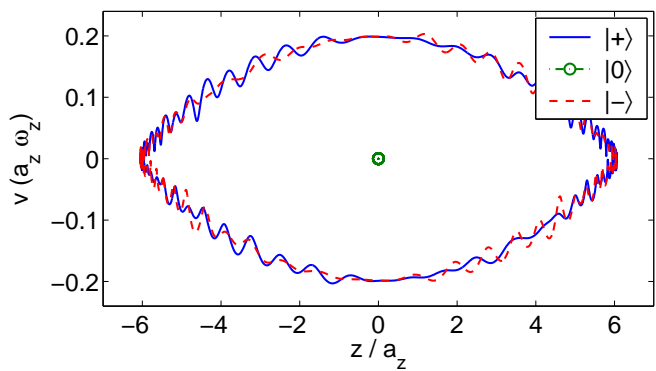}
\caption{(Color online) An eye shaped diagram for the phase portrait of the spinor soliton.
The parameters and notations are the same as in Fig. \ref{fig:s}. Note that the
velocity of the $|0\rangle$ component is very close to zero, and there are many
small oscillations for the $|+\rangle$ and $|-\rangle$ components.}
\label{fig:tse} 
\end{center}
\end{figure}

Figures \ref{fig:tszv} and \ref{fig:tse} display the position and the velocity
of the center of mass (COM) for each component of the solitonic condensate
\cite{Huang01}. Complementary to Fig. \ref{fig:ts}, Fig. \ref{fig:tszv} shows
the oscillations for the last period of our simulation, which lasts from $t=0$
to $t=5000$ $(1/\omega_z)$ with the oscillation period being about $313$
($1/\omega_z$). The velocities of the COM for the $|+\rangle$ and the
$|-\rangle$ components exhibit many small oscillation structures while that of
the $|0\rangle$ component is always very close to zero. We note from Fig.
\ref{fig:tszv} that the position of the COM for the $|0\rangle$ component is
always zero, but the position for the $|+\rangle$ and the $|-\rangle$
components oscillates back and forth periodically and is symmetrically located
on the different sides of that of the $|0\rangle$ component. Figure
\ref{fig:tse} illustrates an eye shaped diagram for the phase portrait of the
$|+\rangle$ and the $|-\rangle$ components during the first period. The ``eye"
opens widely with a clear lid, and the $|0\rangle$ component sits at the
center. We note that the fluctuations of the velocities for both the
$|+\rangle$ and the $|-\rangle$ components are small near the center of the
trap and become enhanced around the turning points of trap boundaries.

Now we comment on the issue of dynamical stability of a
ferromagnetically interacting spin-1 Bose condensate. Earlier we
studied in Ref. \cite{Zhang05b}, the dynamics of a
uniformly distributed spin-1 condensate with ferromagnetic
interactions is unstable. While in this study we see little
distortion of the wave function after an extended real time
propagation even in the presence of initial numerical noise.
This clearly indicates the soliton state is dynamically stable.
If we compare the soliton state with the final domain structure
developed from the homogeneous spin-1 condensate, we see some
similarities to their density distributions. These similarities
point to a
possible connection of the trapped solitonic structure we discuss
to the domain structure in the homogeneous case.
On a more general term, domain structure can be interpreted as
a certain restricted kind of multiple solitons.
Previously, Kasamatsu and Tsubota also noticed
these similarities between the domains and the solitons in
a system of two-component atomic condensates \cite{Kasamatsu04}.

We note that the oscillation period is about $2$ seconds, which is
of the order of the experimental lifetime of the condensate
\cite{Chang04}. Thus, it remains challenging to observe the spin
soliton structure we study here. We also observe that the
quantum diffusion time of the spin in the ground state condensate
is only a few trap cycles, usually shorter than period of the
soliton oscillations \cite{Yi03}. Furthermore, our investigation
focuses on excited state solitonic structures where the
application of the single spatial mode approximation may become
questionable \cite{Yi03}. A more general treatment that
potentially addresses these questions seems completely out of reach
for the time being. On the other hand, as we reduced to $N=10^5$,
still a large number of trapped atoms, the period of the soliton
oscillation reduces due to smaller overlap for the $|\pm\rangle$
components to tunnel through. Thus some aspects of these
solitonic structures could become observable with the system
parameters being tuned to optimize the intended signals before the
condensate disappears.

\begin{figure}
\begin{center}
\includegraphics[width=3.45in]{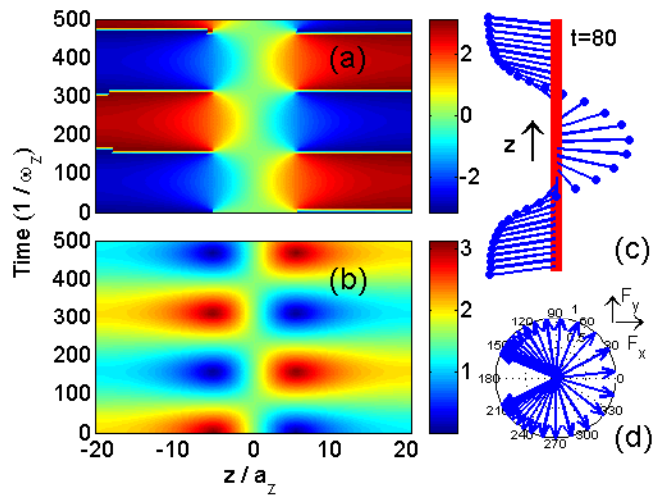}
\caption{(Color online) Spatial and temporal distributions
of the spin evolution for the spinor condensate soliton.
The magnitude of the local spin average is always close to
unity due to the ferromagnetic interaction, but the local spin
direction evolves with time due to the spin exchange interaction.
(a) shows the time dependence of the
spatial distribution for the azimuthal spin angle $\varphi$ and (b)
for the polar spin angle $\theta$, (c) shows the typical spatial
distribution of the spin average (t=80), and (d) shows the projection
of the local spin distributions onto the $x$-$y$ plane.}
\label{fig:ss} 
\end{center}
\end{figure}

The detailed local spin evolution of the soliton state is further
illustrated in Fig. \ref{fig:ss}. The magnitude of the local spin
$f=\sqrt{f_x^2+f_y^2+f_z^2}$ is nearly the same everywhere in
space and time. We believe $f$ is exactly the same if a more
accurate initial state is used. The spatial distribution of the
pointing angles for the local spin, either the azimuthal one $\varphi$
defined by $\tan\varphi
= f_y / f_x$ in \ref{fig:ss}(a) or the polar one $\theta$ defined
by $\cos\theta = f_z / f$ in \ref{fig:ss}(b), are strongly
periodic. Both angles vary rapidly in the central region and
slowly around the boundaries at every moment (the horizontal
directions in Fig. \ref{fig:ss}(a) and (b)). Near the edge of the
condensate (the vertical direction), the angle $\varphi$
essentially remains the same at different times (note $-\pi$ and $\pi$
represent the same spatial direction), while the polar angle
$\theta$ changes slightly during each period. $\theta$ changes most
rapidly with respect to time at about $z=\pm 6 a_z$ where the
$|+\rangle$ and the $|-\rangle$ components take their highest
densities at $t=0$. In the central region, both angles change
moderately with respect to time.

To further clarify the above analysis for the averaged spin
directions, we present the 3D spin distribution at time $t=0$ and
$t=80$ ($1/\omega_z$) in Fig. \ref{fig:s}(a) and Fig.
\ref{fig:ss}(c) and (d), respectively. As stated previously, the
local spins have nearly zero projections in the $y$-axis at time
$t\simeq 0$, $T/2$, $T$, $\cdots$, when they are basically lying in the
$x$-$z$ plane and winding around the $y$-axis with a winding
number of approximately equal to one [Fig. \ref{fig:s}(a)].
Similarly, $f_z\simeq 0$ at time $t\simeq T/4$, $3T/4$, $5T/4$,
$\cdots$, when the local spins are lying in the $x$-$y$ plane but
winding around the $z$-axis with a similar winding number [Fig.
\ref{fig:ss}(c) and (d)]. We find the winding number is a
topological quantity that remains conserved during the
propagation.

\section{Stationary solitons in a rotating reference frame}
\label{sec:sr}

A propagating soliton with a constant speed is actually stationary
in a special moving frame. A wonderful story about this is that J. S.
Russell once followed the water wave soliton for a couple of
miles. Such impressive stability during propagation is
also manifested for a soliton in a spin-1 condensate. We
find that the density of the soliton becomes time independent with
the following canonical transformation,
\begin{eqnarray}
\phi_a &=& {1\over \sqrt 2}\left[{1\over \sqrt 2} (\phi_++\phi_-)
+\phi_0\right], \nonumber\\
\phi_b &=& {1\over \sqrt 2}(\phi_+-\phi_-),
\label{eq:st}  \\ 
\phi_c &=& {1\over \sqrt 2}\left[{1\over \sqrt 2}(\phi_++\phi_-)-\phi_0\right]
\nonumber .
\end{eqnarray}

\begin{figure}
\begin{center}
\includegraphics[width=3.25in]{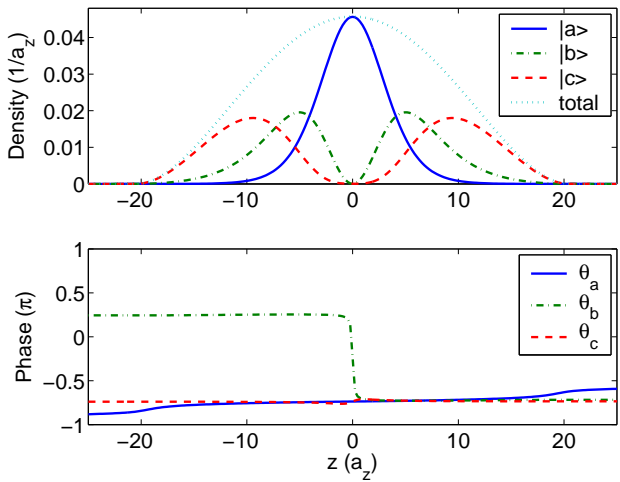}
\caption{(Color online) Density and phase distributions of the soliton after
the transformation (Eq. \ref{eq:st}). The density becomes time independent.}
\label{fig:sn} 
\end{center}
\end{figure}

Figure \ref{fig:sn} shows the density and phase distributions of components
$|a\rangle$, $|b\rangle$, and $|c\rangle$, which are all time independent
after the transformation. The spatial pattern of the phase for each component
is also time independent, i.e., $\theta_a$ and $\theta_c$ are even functions
along the axial coordinate with respect to the center of the condensate,
and $\theta_b$ has a $\pi$ phase shift at the center of the
condensate. We note this density distribution is similar to that of a
Mermin-Ho vortex state of Fig. 1 in Ref. \cite{Mizushima02} again a potential
connection of the solitons and the vortex state.

\begin{figure}
\begin{center}
\includegraphics[width=3.25in]{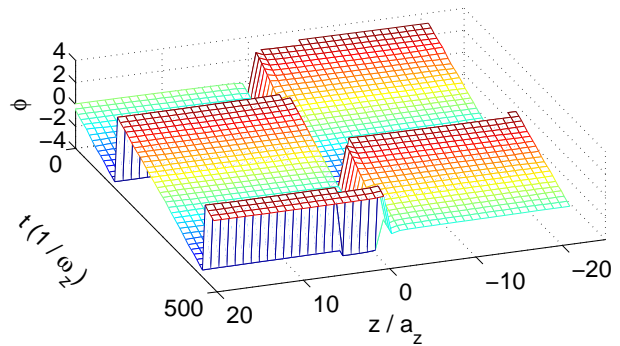}
\caption{(Color online) Azimuthal angle of the soliton after the
transformation. The spatial pattern of the azimuthal angle is time independent
in a rotating reference frame.}
\label{fig:ssn} 
\end{center}
\end{figure}

We can choose a new quantization axis $z'$, different from the
axial direction of the condensate, such that $\phi_a$, $\phi_b$,
and $\phi_c$ correspond to the wave functions of the $|+\rangle$,
$|0\rangle$, and $|-\rangle$ components in the new coordinate
system. We then find the spatial distribution of
$f_{z'}=|\phi_a|^2-|\phi_c|^2$ is time independent (Fig.
\ref{fig:sn}), and the $f_{x'}$ and $f_{y'}$ rotates with a
constant angular velocity around the $z'$ axis as shown in Fig.
\ref{fig:ssn}. If we go further into the rotating frame defined by
\begin{eqnarray}
{\hat x''} &=& {\hat x'} \cos\Omega t + {\hat y'}\sin\Omega t,
\nonumber
\\
{\hat y''} &=& -{\hat x'} \sin\Omega t + {\hat y'}\cos\Omega t,
\\
{\hat z''} &=& {\hat z'}, \nonumber
\end{eqnarray}
where $\Omega = 2\pi/T$, we find the spatial distribution of the
spin of the condensate, which is similar to Fig. \ref{fig:s}(a),
is completely time independent. Thus we show again in the rotating
frame (the double primed one) that the appropriate winding number
is time independent and remains conserved with change of the
reference frame as we confirm previously at the end
of the Sec. \ref{sec:tp}.

\begin{figure}
\begin{center}
\includegraphics[width=3.25in]{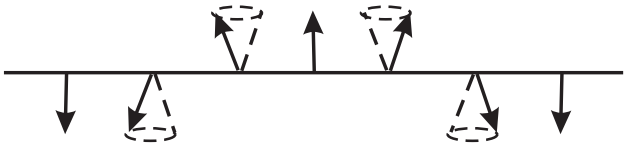}
\caption{Schematic diagram of the dynamics of local spins in the new coordinate system.}
\label{fig:os} 
\end{center}
\end{figure}

Further examination shows that the unitary transformation matrix of Eq.
(\ref{eq:st}) is easily decomposed into spin rotations
\begin{eqnarray}
U &=& \left(\begin{array}{rrr}{1\over 2} & {1\over \sqrt 2} & {1\over 2} \\
     {1\over \sqrt 2} &0 &-{1\over \sqrt 2}\\
    {1\over 2} & -{1\over \sqrt 2} &{1\over 2} \end{array}\right) \nonumber \\
  &=& \exp(i\pi) \exp(-iF_y\cdot {\pi\over 2}) \exp(-iF_z\cdot \pi).
\end{eqnarray}
The new coordinate $(x',y',z')$ after the rotation corresponds to the original
$(z,-y,x)$. Thus $f_{x'} = f_z$, $f_{y'}=-f_y$, and $f_{z'}=f_x$. We have
checked and confirmed in the original coordinate that $f_x$ is indeed
time-independent, i.e., $f_{z'}$ does not change with time (see Fig.
\ref{fig:sn}). We have also found that the time dependence of
$\tan^{-1}(-f_y/f_z)=\tan^{-1}(f_{y'}/f_{x'})=\phi$ is essentially linear as in
Fig. \ref{fig:ssn}. Thus we reach the following simplified picture as Fig.
\ref{fig:os} in the original coordinate with the given initial conditions as
shown in Fig. \ref{fig:s}(a): local spins rotate around the $x$-axis with the
same angular frequency $2\pi/T$. This rotation is constrained so that the
$x$-component of the local spin instead of the usual $z$-component is conserved
in the discussed soliton state. An amazing aspect of this local spin processing
dynamics is that the equivalent gyro-frequency is independent of space.

\section{Conclusion}
\label{sec:cl}

In conclusion, we search numerically for a collectively excited, or a soliton
state in a trapped spin-1 $^{87}$Rb condensate with ferromagnetic
interactions by employing the quasi-one-dimensional nonpolynomial Schr\"odinger
equation (NPSE). We investigate the dynamics of the density, the center of mass
motion, and the local spin distribution of the solitonic state we find
based on a real time
propagation of the NPSE inside the trapped environment.
The tunnelling process of the $|+\rangle$ and the $|-\rangle$
components with the assistance of the $|0\rangle$ component is explained as
mutual precessions among the different components. We further show that the soliton
state is dynamically stable and is related to the domain structure observed
recently in spin-1 $^{87}$Rb condensates \cite{Chang05, Zhang05b}. By transforming
to a rotating reference frame, we demonstrate that the time-dependent dynamics of the
local spin of the soliton state become time-independent. In all of the
reported numerical simulations
we use $N=10^6$ atoms (although unreported simulations were also performed
for $N=10^5$ as further checks). Within the mean field theory we find
that the results are basically the same except that the
period becomes shorter for smaller $N$ due to the reduced overlapped region
for the $|\pm\rangle$ components to tunnel through.

\section{Acknowledgement}
\label{sec:ac}

W. Zhang is indebted to Dr. D. L. Zhou for many stimulating
discussions, and he is also grateful to Dr. J. Ieda for providing more recent
references \cite{Ieda04}. This work is supported by NSF and NASA.
O.E.M. acknowledges support from a TUBA-GEBIP Award.

\end{document}